\documentclass[twocolumn,letterpaper,aps,prc,superscriptaddress,showpacs,nofootinbib,floatfix,amsmath,amssymb]{revtex4-1}

\usepackage{graphicx}% Include figure files
\usepackage{dcolumn}% Align table columns on decimal point
\usepackage{bm}% bold math
\usepackage{hyperref}% add hypertext capabilities
\usepackage{color}
\usepackage[mathlines]{lineno}% Enable numbering of text and display math
\usepackage[normalem]{ulem}
\usepackage{xspace}	% Include xspace

\newcommand{\pt}{\mbox{$p_T$}\xspace}
\newcommand{\pp}{\mbox{$p$$+$$p$}\xspace}
\newcommand{\NN}{\mbox{$N$$+$$N$}\xspace}
\newcommand{\pa}{\mbox{$p$$+$A}\xspace}
\newcommand{\pau}{\mbox{$p$$+$Au}\xspace}
\newcommand{\ppb}{\mbox{$p$$+$Pb}\xspace}
\newcommand{\dau}{\mbox{$d$$+$Au}\xspace}
\newcommand{\heau}{\mbox{$^3$He$+$Au}\xspace}
\newcommand{\sqsn}{\mbox{$\sqrt{s_{_{NN}}}$}\xspace}
\newcommand{\sqsntwo}{\mbox{$\sqrt{s_{_{NN}}}=200$~GeV}\xspace}
\newcommand{\xp}{\mbox{$x_{p}$}\xspace}

\newcommand{\Ncoll}{\mbox{$N_\mathrm{coll}$}\xspace}

\newcommand{\rcp}{\mbox{$R_{CP}$}\xspace}

\newcommand{\rpa}{\mbox{$R_{p+\mathrm{A}}$}\xspace}
\newcommand{\tpa}{\mbox{$T_{p+\mathrm{A}}$}\xspace}
\newcommand{\rdau}{\mbox{$R_{d+\mathrm{Au}}$}\xspace}
\newcommand{\signn}{\mbox{$\sigma_{_{NN}}$}\xspace}

\newcommand{\paper}{paper\xspace}

\newcommand{\colorado}{University of Colorado, Boulder, Colorado 80309, USA}
\newcommand{\bnlphys}{Physics Department, Brookhaven National Laboratory, Upton, New York 11973-5000, USA}

\begin{document}

% DVP - old title for posterity
%\title{Investigating Proton Size Fluctuations with Small Collision Systems at RHIC}

\title{Consequences of high-$x$ proton size fluctuations in small                
  collision systems at RHIC}

% no reason not to have these in alphabetical order? --DVP
\author{D.~McGlinchey} \affiliation{\colorado}
\author{J.~L.~Nagle} \affiliation{\colorado}
\author{D.~V.~Perepelitsa} \affiliation{\bnlphys}

\date{\today}% It is always \today, today,
             %  but any date may be explicitly specified

\begin{abstract}

Recent measurements of jet production rates at large transverse
momentum (\pt) in the collisions of small projectiles with large
nuclei at RHIC and the LHC indicate that they have an unexpected
relationship with estimates of the collision centrality. One
compelling interpretation of the data is that it captures an
$x_p$-dependent decrease in the average interaction strength of the
nucleon in the projectile undergoing a hard scattering. 
A weakly interacting or ``shrinking'' nucleon in the
projectile strikes fewer nucleons in the nucleus, resulting in a
particular pattern of centrality-dependent modifications to high-\pt
processes. We describe a simple one-parameter geometric implementation
of this picture within a modified Monte Carlo Glauber model tuned to \dau jet data, and
explore two of its major consequences. First, the model predicts a
particular projectile-species dependence to the centrality dependence
at high-$x_p$, opposite to that expected from an energy loss effect.
Second, we
find that some of the large centrality dependence observed for
forward di-hadron production in \dau collisions at RHIC 
may arise from the physics of the ``shrinking'' projectile nucleon,
in addition to impact parameter-dependent shadowing or saturation effects at
low nuclear-$x$.
We conclude that analogous measurements in recently collected \pau and \heau collision data at RHIC can provide a unique test of these predictions.

\end{abstract}

% PACS, the Physics and Astronomy
                             % Classification Scheme.
%\keywords{Suggested keywords}%Use showkeys class option if keyword
                              %display desired
\maketitle

\section{Introduction}
\label{sec:intro}

Recent measurements of jet production at large transverse momentum
(\pt) in the collisions of small projectiles (protons and deuterons)
with large target nuclei have revealed an unexpected relationship
between the jet rate and soft-particle production signatures
understood to be related to the collision geometry. The rate of
inclusive jet production was analyzed in $\sqsn = 200$~GeV
deuteron--gold (\dau)~\cite{Adare:2015gla} and $\sqsn = 5.02$~TeV
proton--lead (\ppb)~\cite{ATLAS:2014cpa} collisions at the
Relativistic Heavy Ion Collider (RHIC) and Large Hadron Collider
(LHC), respectively, as a function of the collision {\em centrality},
an experimental handle on the geometric configuration of the
projectile--nucleus collision system.

In these measurements, the rate of jet production in
projectile--nucleus (generically, \pa) collisions was compared to
the expectation derived by scaling the jet rate in individual
nucleon--nucleon (\NN) collisions by the additional degree of nuclear
overlap. The deviation from this expectation
is traditionally quantified with a nuclear modification factor,
\begin{equation}
\rpa = \left.\left( dN^{p+\mathrm{A}}/d\pt \right) \right/
\left( \tpa d\sigma^{p+p}/d\pt \right).
\label{eq:rpa}
\end{equation}

In Eq.~\ref{eq:rpa}, the numerator is the per-event yield in \pa
collisions, while the denominator is the jet production cross-section
in \pp collisions scaled by the nuclear overlap in \pa
collisions, \tpa. Schematically, the denominator may also be understood as
the product of the per-\NN collision yield and \Ncoll, the mean number
of \NN collisions in a projectile--nucleus collision, giving $\tpa           
d\sigma^{p+p}/d\pt = \Ncoll (\sigma_{_{NN}}^{-1}d\sigma^{p+p}/d\pt)$ where
\signn is the inelastic \NN cross-section. \rpa may be
evaluated for different selections on jet kinematics and, as described
below, for different selections on \pa event classes.

When measured in minimum bias (MB)---i.e., centrality-integrated---collisions,
the total rate of jet production was found to be comparable to this
expectation ($\rpa \approx 1$), and any small deviations from this
were generally in line with global analyses of the modifications to
the parton densities in nuclei~\cite{Eskola:2009uj}, and with minimal
energy loss in the initial stages of the
collision~\cite{Kang:2015mta}.

On the other hand, jet rates were also evaluated under different
selections on the event centrality, which in both measurements was
characterized using the total particle activity at large
pseudorapidity in the nucleus-going direction. A Glauber
model~\cite{Miller:2007ri} was used to describe the distribution of
\Ncoll values in \pa collision geometries and
phenomenological models~\cite{Bialas:1976ed} of the relationship
between \Ncoll and the centrality signal were used to determine the
mean \Ncoll in each centrality
selection~\cite{Adare:2013nff,Aad:2015zza}. 

\begin{figure}
        \centering
        \includegraphics[width=0.98\columnwidth]{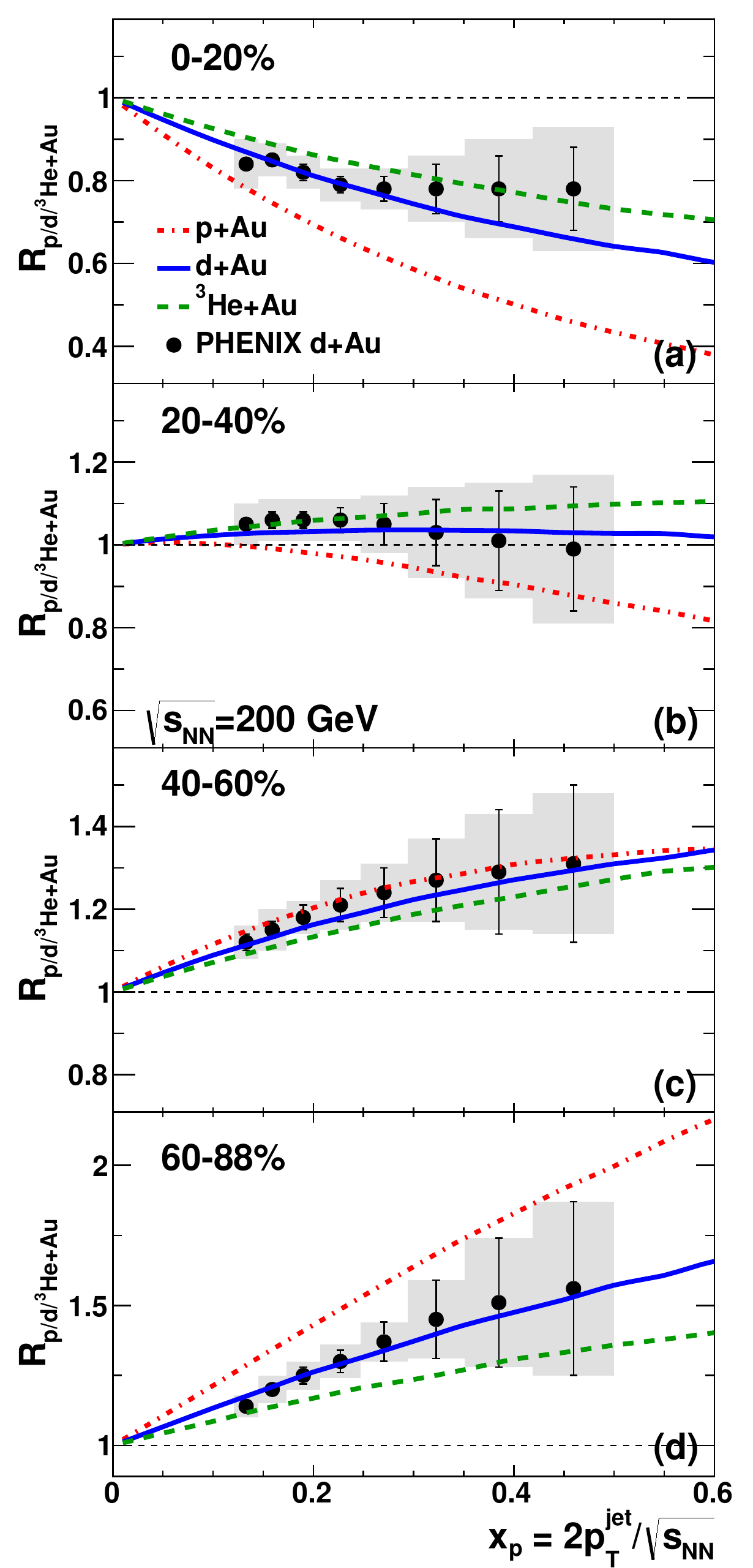}
        \caption{(Color Online) The calculated \rpa as a function of
          \xp in each centrality bin for $p$/$d$/$^3$He$+$Au compared to the measured
          \rdau of jets in \dau collisions at
          \sqsntwo~\cite{Adare:2015gla}.}
        \label{fig:raa}
\end{figure}

Figure~\ref{fig:raa} shows the measured \rpa for \dau collisions (\rdau) in each of the reported centrality ranges. In central collisions,
those associated with small centrality values (0--20\%) and large \Ncoll values,
the jet rate was found to be systematically lower than this
expectation ($\rpa < 1$). In peripheral collisions, those
associated with large centrality values (60--88\%) and small \Ncoll values, the
jet rate was instead systematically higher than the expectation ($\rpa           
> 1$). In both selections, the scale of the deviations from $\rpa =            
1$ generally increased with jet \pt. While the behavior in central
events may be seen as consistent with a modest jet energy loss in the
initial or final states~\cite{Kang:2015mta,Zhang:2013oca}, the
enhanced rate of jet production in peripheral events is
difficult to understand within traditional energy loss
pictures. Furthermore, the centrality-dependent modifications are much
larger than what could be expected from an impact parameter dependence
of the nuclear parton densities~\cite{Helenius:2012wd}. Finally, while
an increase in the soft-particle production rate for \NN collisions with a
hard-scattering is expected to bias centrality-dependent measurements
of jet production, measurements of this correlation in \pp
collisions~\cite{Adare:2013nff,Aad:2015ziq} and estimates of their
impact in \pa
collisions~\cite{Adare:2013nff,Perepelitsa:2014yta,Adam:2014qja}
demonstrate that this effect is small and in the opposite
direction of the observed modifications.

The magnitude of these modifications can be also explored through the
ratio of the \Ncoll-scaled per-event yield between central and
peripheral \pa collisions,
\begin{equation}
%\rcp = \frac{1/N_\mathrm{coll}^\mathrm{central}
%  dN^\mathrm{central}/d\pt}{1/N_\mathrm{coll}^\mathrm{peripheral}
%  dN^\mathrm{peripheral}/d\pt}.
\small
\rcp = \frac{1/N_\mathrm{coll}^\mathrm{central}\
  dN^\mathrm{central}/d\pt}{1/N_\mathrm{coll}^\mathrm{peripheral}\
  dN^\mathrm{peripheral}/d\pt} = \frac{R_{p+\mathrm{A}}^\mathrm{central}}{R_{p+\mathrm{A}}^\mathrm{peripheral}}.
\end{equation}
While the \rpa is necessary to understand the absolute modifications
with respect to the expectation from \NN collisions, the smaller
experimental systematic uncertainties associated with an \rcp measurement allow it
 to quantify the relative modification between two \pa
event classes more precisely.

A unifying way to understand the central, peripheral, and MB data together is to hypothesize that jet
production is unmodified, but the soft-particle production used
to estimate centrality is affected in events with high-\pt
jets. In such an explanation, the overall jet rate is unaffected when
integrated among all types of \pa collisions and is merely
redistributed among the centrality classes, naturally explaining the
observed modifications. Furthermore, the modifications in the data appear only to depend on the longitudinal momentum of the hard-scattered parton in the projectile, \xp, and on no other kinematic variable. While the \rpa modifications at RHIC and the LHC appear at very different $p_{T}^\mathrm{jet}$ ranges, they are at similar ranges in \xp. Additionally, the ATLAS \rpa analysis demonstrated that the results at multiple rapidity selections have a universal dependence only on \xp~\cite{ATLAS:2014cpa}.

A compelling underlying description of such an effect is that it arises from color fluctuations in the internal configuration of the projectile nucleon. Since hadrons are composite objects, their Fock space description is a distribution over quark-gluon configurations with varying properties. In hadronic collisions, the different possible internal configurations result in event-by-event fluctuations in the effective interaction cross-section~\cite{Heiselberg:1991is,Blaettel:1993ah}. For the hadronic configurations
with a large-\xp parton available for a hard scattering, the average
cross-section is expected to decrease on general grounds since, for
example, such configurations have a fewer than average number of
partons~\cite{Brodsky:1973kr}. Thus, when passing through a large
nucleus, these weakly interacting or geometrically ``shrinking''
configurations interact with fewer nucleons than an average-sized
configuration, resulting in a relative decrease in the \Ncoll
distribution for these events with a large-$x_p$ scattering parton.

Such an interpretation of the data was first proposed by the
authors of Ref.~\cite{Alvioli:2014eda} (and see references
therein). Other authors have described empirical models in which a
similar effect can be achieved through a depletion of the longitudinal
energy of the projectile remnant after the removal of a high-\xp
parton~\cite{Armesto:2015kwa} or from an overall \xp-dependent
suppression in the soft particle multiplicity per-\NN
collision~\cite{Bzdak:2014rca}. We do not discuss the merits of these
approaches here, but rather focus on the proton color fluctuation picture,
since it has a simple geometric interpretation which can be
implemented in modern Monte Carlo Glauber (MC-Glauber) approaches.

In this \paper, we argue that a systematic scan of
centrality-dependent hard process rates in small collision systems at
RHIC as a function of the projectile nucleus mass number (A=1,2,3) can
better explore the underlying physics mechanisms. Specifically, this
can be performed with measurements of high-\pt jet or particle
production rates in high-luminosity \pau,
\dau, and helium--gold (\heau) collision data taken
in 2014, 2008, and 2015, respectively. On general grounds, a
suppression of the \rpa in central \pa collisions which
arises from energy loss should become stronger with increasing
projectile mass number, since the amount of nuclear material and thus
path length increases. On the other hand, a suppression of the central
\rpa (and enhancement of the peripheral \rpa) which arises from the
shrinking of a high-\xp hard-scattered projectile nucleon should
generally decrease with increasing projectile mass number, since the
\Ncoll contributed by the other nucleons in the projectile which did
not undergo the hard scattering would be unaffected. Thus, the unique
capability of RHIC to deliver $p/d/^{3}$He+Au collisions at the same
energy can serve as a novel way to constrain the underlying
phenomenon.

In addition, we argue that the effects of proton color fluctuations at
large projectile-\xp contributes to the observed centrality
dependence of forward hadron and di-hadron production rates in \dau
collisions at RHIC~\cite{Adler:2004eh,Adare:2011sc,Abelev:2007nt,Arsene:2004ux}. The strong centrality dependence observed
in these collisions has been previously taken to be evidence of
non-linear QCD effects at small nuclear-$x$. 
%However, centrality
%classes in \dau collisions select only broad distributions over \Ncoll
%or impact parameter. Additionally, 
However, high-\pt particle production in this
kinematic regime is dominated by $\xp >               
0.1$, meaning that proton color fluctuations affect the apparent centrality dependence of hard scattering rates. We
demonstrate that this effect should be accounted for in order to quantify 
the modifications from remaining nuclear effects.

Our \paper is organized as follows. In Section~\ref{sec:model}, we
implement a one-parameter model of an \xp-dependent decrease in the
\NN interaction strength in a MC-Glauber approach and
describe its application to the centrality frameworks used by the
PHENIX Collaboration. In Section~\ref{sec:results}, we give the
predictions of our model, after tunning to recent \dau measurements, for
the \rpa and \rcp in \pau and \heau collisions. In
Section~\ref{sec:dihadron} we discuss the consequences of this picture
for interpretations of forward di-hadron production measurements at
RHIC. We finish with a discussion and summary in
Section~\ref{sec:summary}.

\section{Model}
\label{sec:model}

This section describes a simple geometric interpretation of an \xp-dependence of the nucleon-nucleon interaction strength for the hard scattering projectile nucleon within a Glauber model picture of \pa collisions. The model is discussed in Sec.~\ref{sec:signnmod}, and its implementation in a MC-Glauber simulation and application to the centrality frameworks used by experiments is detailed in Sec.~\ref{sec:glauber}. 

\subsection{Geometric picture of shrinking proton}
\label{sec:signnmod}

\begin{figure}
	\centering
	\includegraphics[trim={0 3cm 0 0 }, clip, width=0.9\columnwidth]{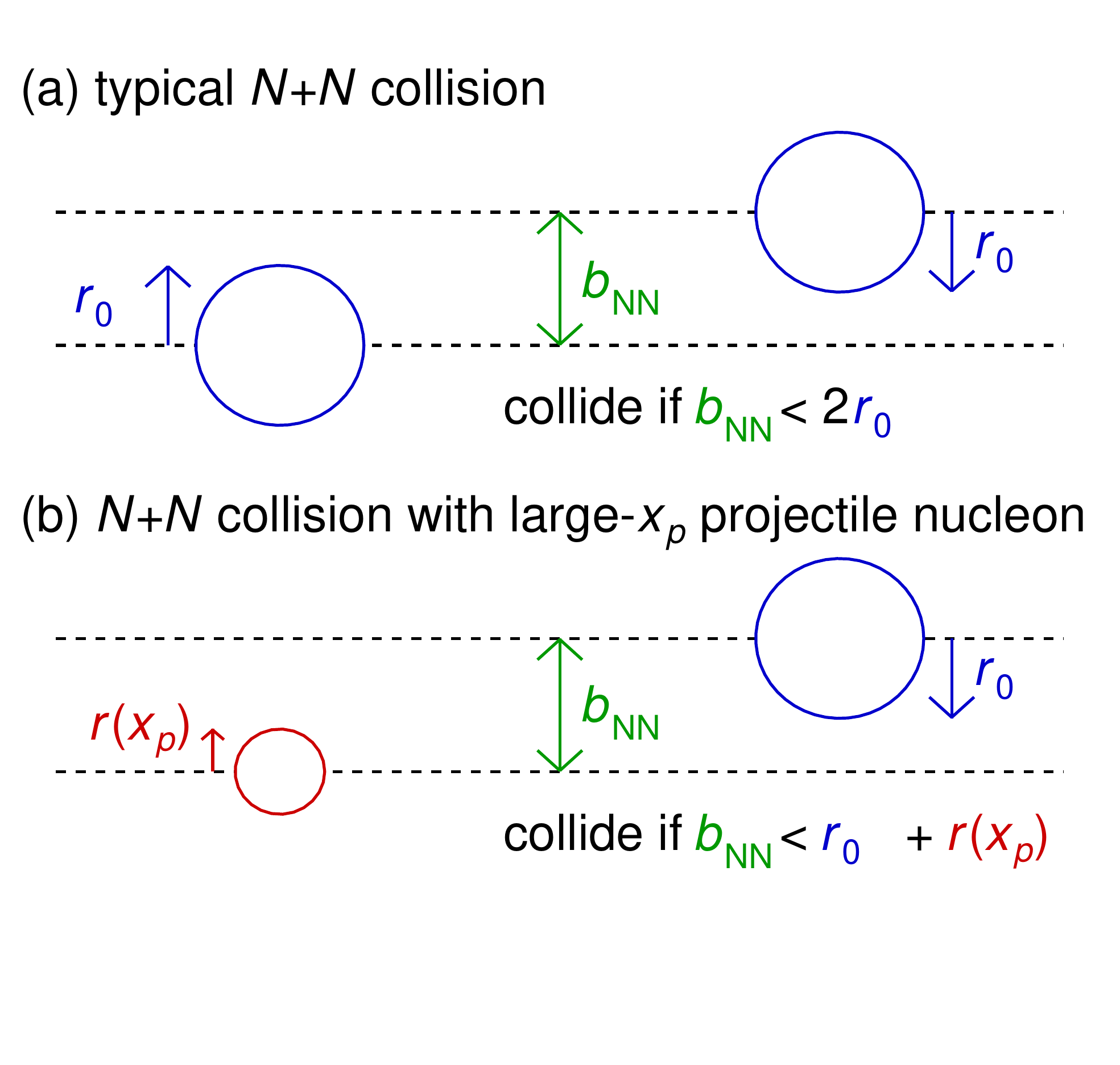}
	\caption{(Color Online) Schematic picture of a nucleon-nucleon collision for the case of (a) an unmodified projectile nucleon and (b) a projectile nucleon whose radius depends on \xp.}
	\label{fig:schematic}
\end{figure}

Our analysis models the hypothesis that in a high-energy \pa collision in which a large-\xp parton in the projectile undergoes a hard-scattering, the nucleon--nucleon interaction cross-section \signn between the projectile nucleon and nucleons in the nucleus is reduced. Classically, the collision cross-section $\sigma$ for two hard spheres with radii $r_1$ and $r_2$ is given by $\sigma = \pi(r_1 + r_2)^{2}$. By analogy, for a projectile nucleon with a large-\xp hard-scattered parton, let the interaction radius, $r$, be dependent on the $x$ of the projectile parton, \xp, and given by $r(\xp)$. The interaction radius for all other nucleons is given by $r_0 = \sqrt{\sigma_{_{NN}}/4\pi}$. A schematic view of colliding nucleons in the modified and unmodified case is shown in Fig.~\ref{fig:schematic}. As argued above, at large \xp ($\gtrsim 0.1$), $r(\xp) < r_0$ and decreases with increasing \xp. We model $r(\xp)$ phenomenologically as a decreasing exponential function characterized by one parameter $\beta$,
\begin{equation}
r(\xp) \equiv \exp(-{\beta}{\xp}) r_0.
\end{equation}

This functional form decreases smoothly and continuously with \xp without introducing many parameters. The resulting nucleon--nucleon cross-section, $\sigma(\xp)$, is therefore \xp-dependent and is given by
\begin{equation}
\sigma(\xp) \equiv \pi(r_0 + r(\xp))^{2} = \frac{1}{4}\left(1 + \exp(-\beta{\xp})\right)^{2}\sigma_{_{NN}}.
\label{eq:sigmod}
\end{equation}

In our treatment, the high-\xp configuration is taken to be frozen throughout the duration of the collision. Furthermore, the interaction radius of every other nucleon in the \pa system is unmodified. Thus, the interaction cross-section between any pair of nucleons which do not contain the large-\xp projectile nucleon is \signn. 

%While the \xp-dependent decrease in the interaction strength may well depend on collision energy, in this work we consider only RHIC energies and therefore maintain a fixed $\beta$. 

\begin{figure}
	\centering
	\includegraphics[width=0.9\columnwidth]{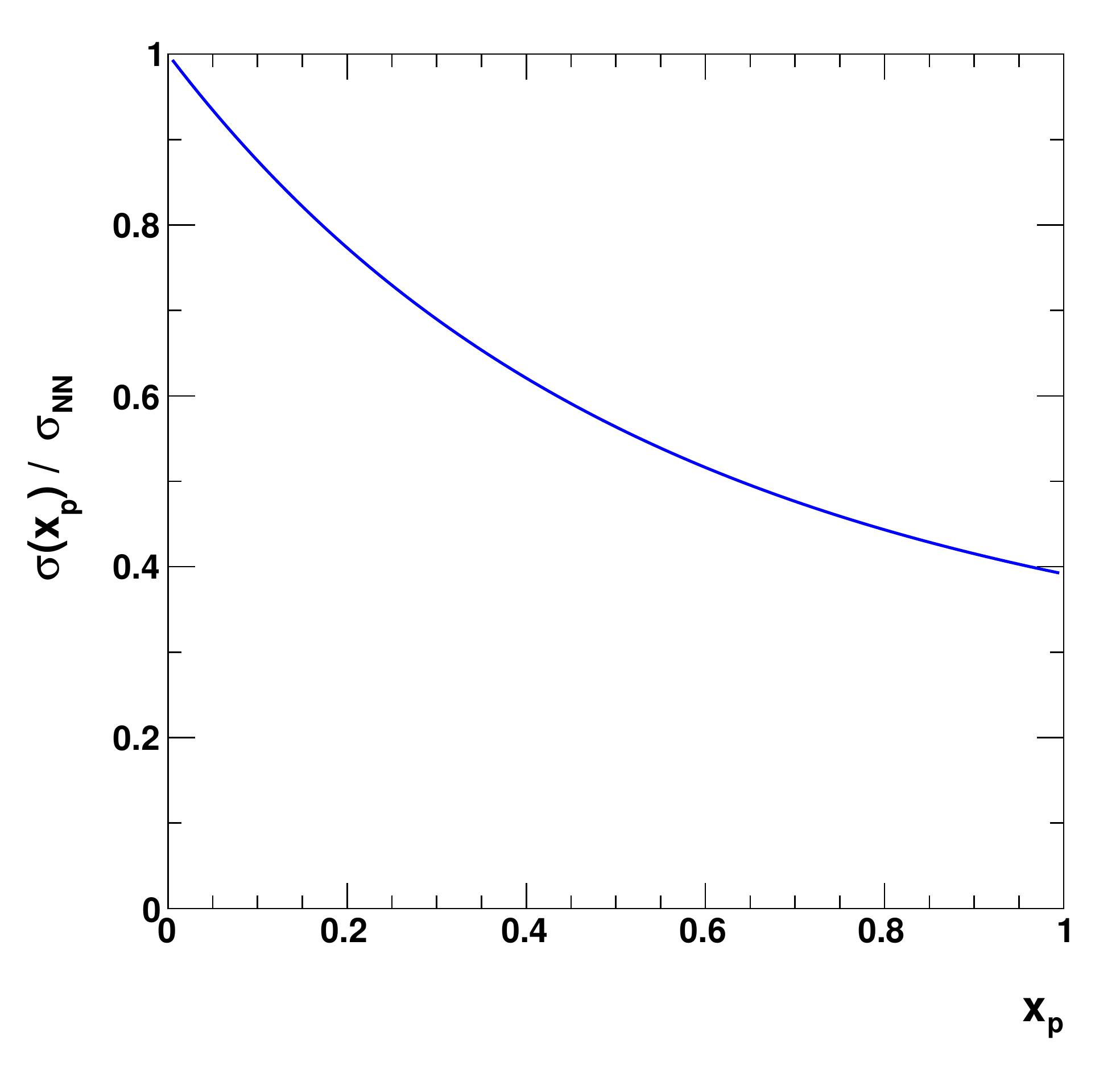}
	\caption{Ratio of the modified nucleon-nucleon cross section $\sigma(\xp)$ to the nominal value \signn, as a function of the momentum fraction in the projectile nucleon (\xp).}
	\label{fig:sigmod}
\end{figure}

In our analysis, $\beta$ was ultimately determined to be $1.38^{+0.09}_{-0.07}$ by implementing this shrinking nucleon picture in a Glauber Monte Carlo analysis, convolving with the centrality framework used by PHENIX, and fitting to the \dau centrality-dependent jet data in Ref.~\cite{Adare:2015gla}. These steps are described in detail in the next sections. The resulting $x_{p}$-dependent cross-section as a fraction of the nominal \signn is shown in Fig.~\ref{fig:sigmod}. Schematically, the value of $\beta = 1.38$ implies that for projectile nucleon configurations with $\xp \sim 0.2$ ($\sim 0.6)$, $\sigma(\xp)$ is reduced by $\sim25$\% ($\sim50$\%) relative to \signn. 

\subsection{Application to experimental centrality models}
\label{sec:glauber}

To determine the effects of the shrinking proton on experimentally measured centrality-dependent hard-scattered yields, the consequences of Eq.~\ref{eq:sigmod} are implemented in the MC-Glauber analysis and centrality framework used by the PHENIX experiment in Ref.~\cite{Adare:2013nff}. The PHOBOS Glauber Monte Carlo code~\cite{Loizides:2014vua} is used to simulate the possible range of $p$/$d$/$^{3}$He+Au collision geometries. In the MC-Glauber approach, the transverse positions of the nucleons in both the target and projectile nuclei are sampled on an event-by-event basis, with each nucleon required to have a minimum separation distance of 0.4~fm from all others. We use the default prescriptions present in the PHOBOS MC-Glauber for describing the radial distributions of nuclei in $^3$He and Au nuclei. To model the deuteron, the Hulth\'{e}n form is chosen~\cite{Hodgson}. 

In all three collision systems, the nucleon-nucleon cross section \signn is chosen to be $42$~mb at the center-of-mass energy $\sqrt{s} = 200$~GeV. For each simulated event, the projectile nuclei are displaced in the transverse plane by a random impact parameter, and \Ncoll is determined by projecting nucleons along straight line (longitudinal) trajectories. Any projectile-target nucleon pair is treated as having interacted if their transverse impact parameter is smaller than the sum of the interaction radii. For unmodified \NN interactions, this condition is $b_{_{NN}} < 2r_0 = \sqrt{\sigma_{_{NN}}/\pi}$. After many simulated events, a given collision system can be described by a probability distribution over the number of \NN collisions, $P(\Ncoll)$.

%To simulate events in which one of the nucleons in the projectile participated in a large-$x_{p}$ hard scattering, the cross-section for all nucleon--nucleon interactions in which that projectile nucleon participates is modified to be $\sigma(x_{p})$. In \dau and \heau collisions, the projectile nucleon is chosen at random in each simulated event. The remaining nucleons are unmodified. Thus, the MC-Glauber simulation takes an additional input parameter, $x_{p}$, and the set of $p$+A collisions generated with a given $x_{p}$ generally have different \Ncoll distributions whose mean \Ncoll decreases systematically with increasing $x_{p}$ and decreasing $\sigma(x_{p})$.

In the PHENIX experiment, classes of $p$+A collisions are selected indirectly through the total charge, $Q$, measured in the beam--beam counter (BBC), situated at $-3.9 < \eta < -3.0$, i.e. in the gold-going direction.
%downstream of the gold nucleus~\footnote{In this work, $\eta < 0$ and $\eta > 0$ are taken by convention to be the directions downstream of gold and projectile nuclei, respectively.}. 
A given range of $Q$ values is related to a set of $p$+A collisions by assuming that all collisions with a given value of \Ncoll produce a distribution over the charge given by a negative binomial distribution (NBD) with \Ncoll dependent parameters. A NBD over $Q$ can be described by a mean $\mu$ and positive exponent $\kappa$, with probability mass function defined by\footnote{This definition differs from the one in Ref.~\cite{Adare:2013nff} only in the $(1 +\mu/\kappa)^{-\kappa}$ normalization term, which in this paper ensures that $\sum_{Q} NBD(Q) = 1$ explicitly.}
\begin{equation}
	\small
	NBD(Q;\mu,\kappa) = \left(1+\frac{\mu}{\kappa}\right)^{-\kappa}\frac{(\kappa+Q-1)!}{Q!(\kappa-1)!}\left(\frac{\mu}{\mu+\kappa}\right)^Q.
	\label{eq:NBD}
\end{equation}
The parameters are taken to grow linearly in \Ncoll, $\mu(\Ncoll) = \mu \Ncoll$ and $\kappa(\Ncoll) = \kappa \Ncoll$. That is, the distribution at fixed \Ncoll is an \Ncoll-fold convolution of an elemental distribution with parameters $\mu$ and $\kappa$. 

Since the MB trigger condition is not fully efficient for peripheral $p$+A events, the efficiency $\epsilon_{MB}$ at a given value of $Q$ is modeled with the following two-parameter functional form,
\begin{equation}
\epsilon_{MB}(Q) \equiv 1 - \exp\left( (Q/p_0)^{p_1} \right),
	\label{eq:eff}
\end{equation}
where $p_0$ and $p_1$ are determined from fits to data. Thus, for a given input distribution $P(\Ncoll)$, the resulting per-event MB $dN/dQ$ distribution measured in PHENIX is given by
{\small
\begin{align}
\frac{dN}{dQ} \equiv & \frac{\epsilon_{MB}(Q)}{\epsilon_{MB,tot}} \nonumber \\
& \left( \sum_{\Ncoll} P(\Ncoll) NBD(Q; \mu\Ncoll, \kappa\Ncoll) \right),
\label{eq:dNdQ_MB}
\end{align}
}
where $\epsilon_{MB,tot}$ is the MB trigger efficiency for all $p$+A collisions and serves as an overall normalization constant. The resulting distribution is then divided into a number of centiles according to $\epsilon_{MB,tot}$.

The charge distribution $dN^\mathrm{hard}/dQ$ for events with a
generic hard process (one expected to obey \Ncoll-scaling) is
identical to Eq.~\ref{eq:dNdQ_MB} but with an \Ncoll-weighting within
the summation,
{\small
\begin{align}
\frac{dN}{dQ}^\mathrm{hard} \equiv &
\frac{\epsilon_{MB}(Q)}{\epsilon_{MB,tot}} \nonumber\\
& \left( \sum_{\Ncoll} \Ncoll
P(\Ncoll) NBD(Q;  \mu\Ncoll, \kappa\Ncoll) \right).
\label{eq:dNdQ_hard}
\end{align}
}%

Since the performance of the BBC system was found by PHENIX to have a
small run-dependence, the $\mu$ and $\kappa$ parameters are slightly
different for each collision system. These small performance
differences, in addition to the different \Ncoll distributions, also
result in collision-system dependent values of $p_0$, $p_1$ and
$\epsilon_{MB,tot}$. All five parameters are summarized in
Table~\ref{tab:NBDpar}. For brevity, we denote the most peripheral selection for all collision systems as 60--88\%, even though it is really 60--84\% for \pau and \heau collisions.

% DVP: needed to make vertical spacing in Table not look crappy
\renewcommand{\arraystretch}{1.3}

\begin{table}
	\caption{The negative binomial distribution parameters $\mu$ and $\kappa$, trigger efficiency parameters $p_0$ and $p_1$, and total MB trigger efficiency $\epsilon_{MB,tot}$ used to characterize the centrality in each collision system by the PHENIX experiment~\cite{Adare:2013nff,pAuHeAuPHENIXNumbers}.}
	\label{tab:NBDpar}
	\begin{tabular}{cccccc}
	\hline\hline
	System & $\mu$ & $\kappa$ & $p_0$ & $p_1$ & $\epsilon_{MB,tot}$ \\
	\hline
	\pau  & 3.14 & 0.47 & 1.076 & 0.602 & 84\%\\
	\dau  & 3.04 & 0.46 & 0.897 & 0.612 & 88\% \\
	\heau & 2.91 & 0.55 & 1.221 & 0.510 & 84\% \\
	\hline\hline
	\end{tabular}
\end{table}

In this paper, our emphasis is on exploring the projectile-species dependence of shrinking projectile nucleon effects. We found that the simple modification to the PHOBOS MC Glauber simulation described above was sufficient for this purpose. However, other authors have raised the importance of improved modeling of the nucleon-nucleon correlations in the nucleus or of the localization of high-$x$ partons in the transverse core of nucleons~\cite{Alvioli:2009ab}. While including these effects could slightly alter the quantitative results, they would not change our essential conclusions.

\section{Results and discussion}
\label{sec:results}

To determine the nuclear modification factors \rpa and \rcp arising
from proton color fluctuations, the simulated $p$/$d$/$^{3}$He+Au
events are re-analyzed under the hypothesis of a shrinking projectile nucleon. 
%The nucleon--nucleon interaction cross-section for the hard-scattered projectile nucleon to
%undergo soft interactions with the nucleons in the nucleus, which
%generate the centrality signal $Q$, is reduced. This reduction is
%stronger with increasing $x_p$ as given in Eq.~\ref{eq:sigmod}. 
For each simulated $p$/$d$/$^{3}$He+Au event, the number of soft
nucleon--nucleon interactions is recalculated by modifying
\signn for the hard-scattered projectile nucleon to be
$\sigma(\xp)$. In \dau and \heau collisions, the affected projectile nucleon is chosen at random in each simulated event. In each event, this results in a modified number of total
\NN collisions $\Ncoll^\prime$, which is a function of
\xp. Operationally, in Eq.~\ref{eq:dNdQ_hard} the $\Ncoll$-dependent
centrality signal for events with a hard scattering is modified via
\begin{align}
NBD(Q; &\mu\Ncoll, \kappa\Ncoll) \rightarrow \nonumber \\
& NBD(Q; \mu N^\prime_\mathrm{coll}(\xp), \kappa N^\prime_\mathrm{coll}(\xp) ),
\end{align}

\noindent yielding an \xp-dependent $Q$ distribution for
hard-scattered events, $dN^\mathrm{hard}/dQ(\xp)$. This analysis is
repeated over a wide range of $x_p$ values in small steps of
\xp. The resulting \rpa values as a function of \xp are obtained for
each centrality selection (exclusive $Q$ range) by evaluating the
ratio of the integral of the modified $dN^\mathrm{hard}/dQ(\xp)$
distribution to the unmodified one,
\begin{equation}
\rpa(\xp) \equiv \left. \int_{cent} dQ \frac{dN}{dQ}^\mathrm{hard} \!\!\!\!\!(x_p) \right/ \int_{cent} dQ \frac{dN}{dQ}^\mathrm{hard}.
\end{equation}

By construction, for MB (i.e. centrality---or $Q$---integrated)
collisions, the \rpa is unity. The resulting \xp-dependent \rcp
values are ratios of the \rpa values in the
corresponding centrality bins, $\rcp(\xp) = R_{p+\mathrm{A}}^\mathrm{central}(x_p) / R_{p+\mathrm{A}}^\mathrm{peripheral}(x_p)$.

\begin{figure}
        \centering
        \includegraphics[width=0.98\columnwidth]{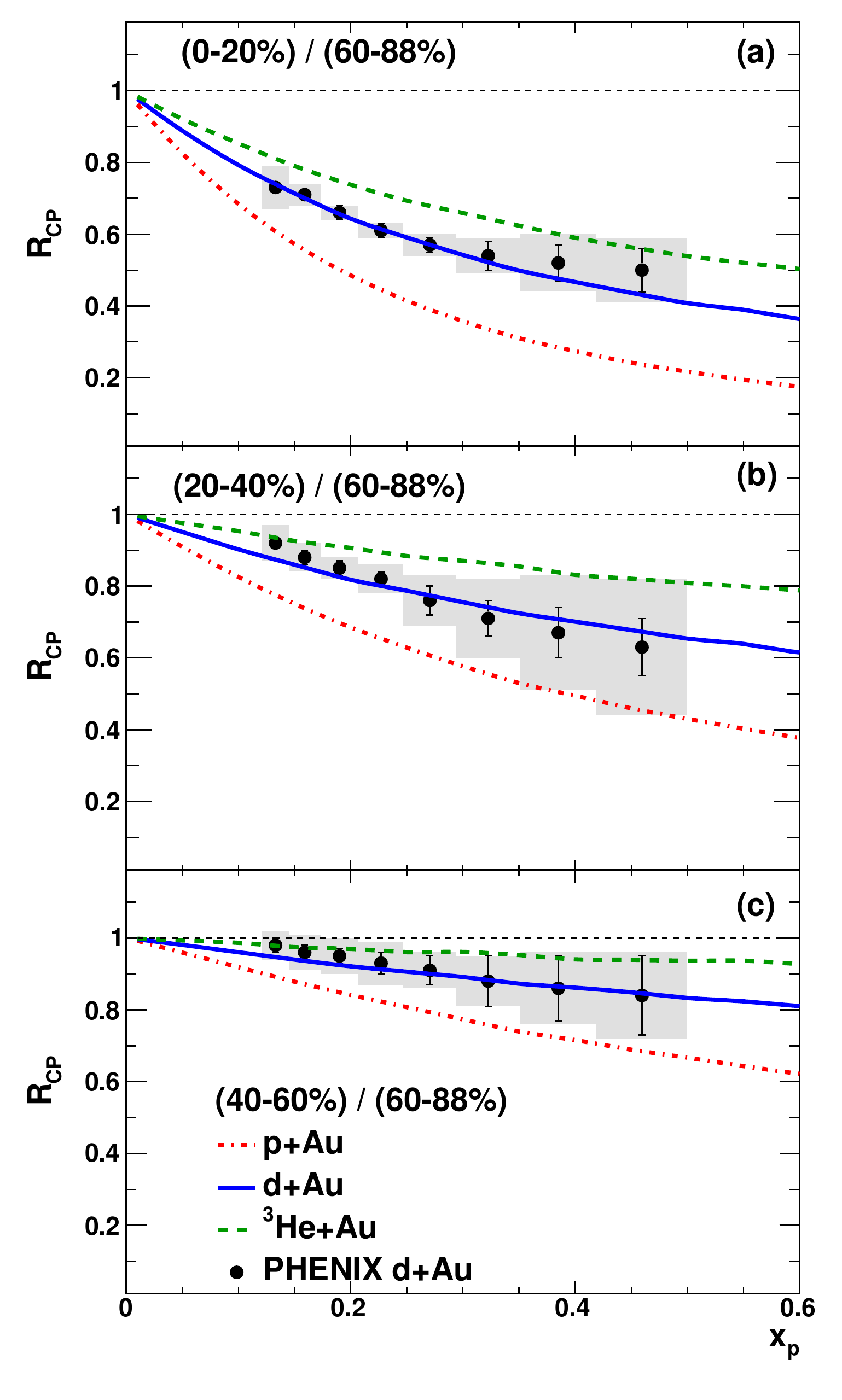}
        \caption{(Color Online) The calculated \rcp as a function
          of \xp in each centrality bin compared to the measured
          \rcp of jets in \dau collisions at
          \sqsntwo~\cite{Adare:2015gla}.}
        \label{fig:rcp}
\end{figure}

Figures~\ref{fig:raa} and~\ref{fig:rcp} show the resulting \rpa and \rcp values
 as a function of $x_p$ for the \pau, \dau and \heau systems. The calculations
are compared to the \dau centrality-dependent jet \rdau and \rcp
measurement by PHENIX~\cite{Adare:2015gla}, under the
assumption that the production of jets at $y=0$ corresponds to typical
\xp values given by
\begin{equation}
\xp(p_T^\mathrm{jet}) = 2p_T^\mathrm{jet} / \sqrt{s_{_{NN}}}.
\end{equation}

The value of $\beta = 1.38^{+0.09}_{-0.07}$ was determined by fitting the calculated
\dau $\rcp$ in the most central-to-peripheral selection (0--20\%/60--88\%) to
that in data as a function of $x_p$.  The figures show only the central value $\beta$ result
as the uncertainty band is small relative to the uncertainty in the data.
%\NOTE{Jamie: Probably need a sentance somehwere saying that we only plot the central value due to the relatively small uncertainty in the $\beta$ -- Darren} 
Our one-parameter model is sufficient to describe the \xp-dependence 
of the measured jet \rpa and \rcp values in \dau collisions.

The same value of $\beta$ is used to calcuate the \rpa and \rcp as a function of
\xp for \pau and \heau collisions at RHIC, where
centrality-dependent hard-process rates have not yet been
reported. A clear ordering between the collision systems is
observed in each centrality selection, with the \rcp for \heau
collisions less suppressed than that for \dau collisions, which is
itself less suppressed than that for \pau collisions. 
The predicted modifications for \pau collisions are
particularly large, showing a 50\% suppression in central collisions
and a factor of two enhancement in peripheral collisions, at $\xp \sim0.4-0.5$. 

This ordering is a direct consequence of the presence of one (two) additional
projectile nucleons in \dau (\heau) collisions compared to the single
projectile nucleon in \pau collisions. The additional projectile
nucleons, which did not undergo a hard-scattering and which do not
generally have a large-$x_p$ parton, have an unmodified interaction
cross--section \signn. Thus, the centrality signal generated by
soft nucleon--nucleon collisions in which these additional projectile
nucleons participate is unmodified. This dilutes the effect that the
shrinking of the hard-scattered projectile nucleon has on the
$dN^\mathrm{hard}/dQ$ distribution and brings the \rpa and \rcp values closer to unity.

%Fig.~\ref{fig:raa} shows the analogous prediction for the
%centrality-selected \rpa as a function of $x_p$ for
%$p$/$d$/$^{3}$He+Au collisions. As discussed above, the
%centrality-integrated \rpa in our model is unity by construction for
%all collision systems. The calculated \rpa for \dau collisions,
%constrained to the centrality-selected \dau \rcp data in only one
%centrality bin, is able to describe the \rdau data in all centrality
%classes, with a suppression in central events and enhancement in
%peripheral ones.  When compared across the three collision systems,
%the calculated \rpa shows a clear hierarchy between them. 

%In the most central collisions, the calculated \rpa for \pau (\dau) collisions
%shows a stronger (weaker) suppression than that for \dau collisions,
%while in the most peripheral collisions it shows a stronger (weaker)
%enhancement. 

%In the two intermediate centrality selections, the
%calculated \rpa values for all three systems are more similar and
%are closer to unity.

Our model predicts that for collisions of a projectile
composed of $N$ nucleons with a large nucleus, the modifications in the \rpa and
\rcp which result from proton color fluctuations will be diluted by a
factor $1/N$ relative to \pa collisions, and that at fixed \xp the following ordering is expected,
\begin{align}
\text{Shrinking}&\text{ nucleon:}\nonumber \\
% R_{CP}^{p+\mathrm{Au}} < R_{CP}^{d+\mathrm{Au}} < R_{CP}^{^{3}\mathrm{He}+\mathrm{Au}}.
& R_{p+\mathrm{Au}}^\mathrm{central} < R_{d+\mathrm{Au}}^\mathrm{central} < R_{^3\mathrm{He}+\mathrm{Au}}^\mathrm{central}.
\label{eq:RCP_ordering}
\end{align}

\noindent An inverted ordering would apply in the most peripheral collisions. On the other hand, if the modifications arise from an effect which
grows with the amount of nuclear material in the collision (such as an
initial- or final-state energy loss of hard-scattered partons in the
nuclear medium), the opposite ordering may be expected,
\begin{align}
\text{Energy}&\text{ loss:}\nonumber \\
% R_{CP}^{p+\mathrm{Au}} > R_{CP}^{d+\mathrm{Au}} > R_{CP}^{^{3}\mathrm{He}+\mathrm{Au}}.
& R_{p+\mathrm{Au}}^\mathrm{central} > R_{d+\mathrm{Au}}^\mathrm{central} > R_{^3\mathrm{He}+\mathrm{Au}}^\mathrm{central}.
\end{align}

These competing descriptions of the data can be tested directly at
RHIC with measurements of centrality-dependent hard-process rates in
the recently-collected \pau and \heau collision data. 
%The predicted
%ordering in Eq.~\ref{eq:RCP_ordering} is a generic consequence of
%increasing the number of nucleons in the projectile, and should hold
%for any reasonable value of $\beta$. To make these predictions
%quantitative, we have constrained $\beta$ in our model using \dau
%collision data and provide quantitative predictions for the analogous
%modifications in \pau and \heau collisions as a function of $x_p$.

% While we have presented calculations for both the full set of
% centrality-dependent \rcp and \rpa values in Figures~\ref{fig:rcp}
% and~\ref{fig:raa}, respectively, we suggest that examining the
% modifications in the most central to most peripheral \rcp can best
% reveal the differences between the three collision systems. This is
% because the expected signature is particularly large in the \rcp and,
% experimentally, an \rcp measurement generally benefits from a
% cancellation of the systematic uncertainties in the ratio of two
% measured spectra and from a reduced overall normalization
% uncertainty. At $\xp=0.3$, a kinematic range which should be reachable
% at mid-rapidity in the recently collected data, the calculated
% 0--20\%/60--88\% \rcp values are $0.38$, $0.57$, and $0.69$ in \pau,
% \dau, and \heau collisions respectively.

\begin{figure}
        \centering
        \includegraphics[width=0.98\columnwidth]{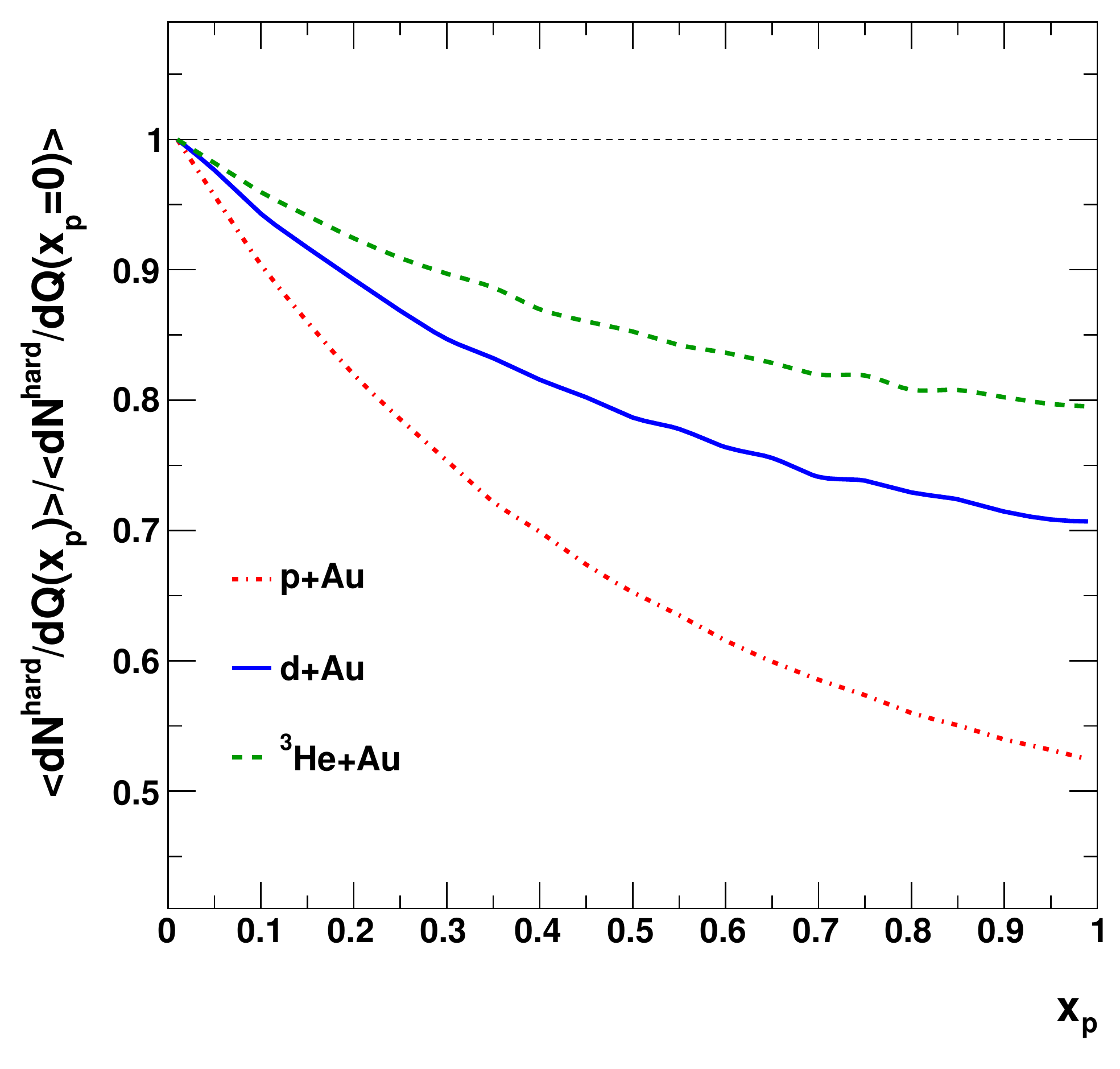}
        \caption{(Color Online) The mean charge in the Au-going BBC
          for each collision system as a function of \xp, normalized
          to the value at $\xp=0$.}
        \label{fig:BBCQ}
\end{figure}

Finally, since the modifications shown in Figs.~\ref{fig:raa} and~\ref{fig:rcp}
arise from an \xp-dependent modification of the centrality signal
(which in PHENIX is the Au-going beam--beam counter charge, $Q$), an alternative but indirect way to explore this signature
is by examining the mean value of $Q$ in
hard-scattered events as a function of \xp. In our model, this is
given directly by the mean value of
$dN^\mathrm{hard}/dQ(\xp)$. Figure~\ref{fig:BBCQ} shows the dependence
of the mean value of $Q$ on \xp for all
three collision systems. Since the overall scale of the mean charge is
very different between the three systems, the results are plotted in ratio to the value at $\xp = 0$.
Mathematically, this value is equal to the mean $Q$ for hard-scattering events in the absence of an \xp-dependent shrinking of the projectile nucleon. 
A clear hierarchy between the
three systems is visible at each value of \xp, with the largest
relative suppression at fixed \xp in \pau collisions, and a
systematically smaller reduction in the mean charge as additional
nucleons are added to the projectile.

\section{Di-hadron correlations}
\label{sec:dihadron}

\begin{figure}
        \centering \includegraphics[width=0.98\columnwidth]{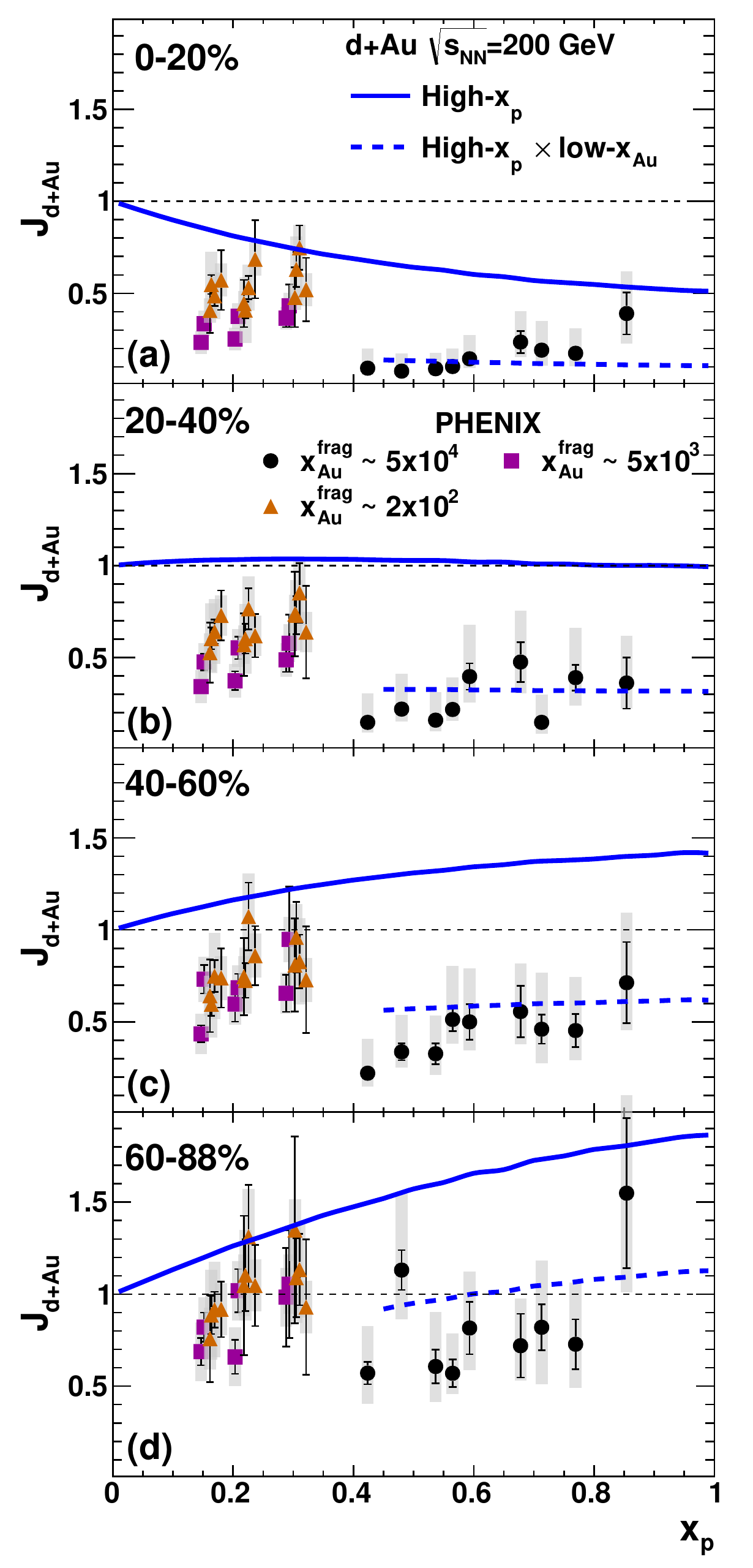}
        \caption{(Color Online) The measured $J_{d+\mathrm{Au}}$ as a function of
          \xp~\cite{Adare:2011sc} in each centrality bin compared to the calculated
          modification due to the shrinking projectile nucleon
          size in \dau collisions. }
        \label{fig:jda}
\end{figure}

The \dau data analyzed in the previous section are from measurements
of jet production at midrapidity which, along with our predictions,
extends to the kinematic region $\xp \sim 0.4-0.5$. In the future, the
large acceptance and hadronic calorimetry of the sPHENIX
experiment~\cite{Adare:2015kwa}, along with the projected performance
for the luminosity of \pau collisions at RHIC, will enable jet
measurements at midrapidity which extend this range to $\xp \approx                   
0.75$. While our model is constrained by data in the region $\xp<0.5$, it may be naturally extended to provide predictions at higher values as given by Eq.~\ref{eq:sigmod}. Figure~\ref{fig:jda} shows that the predicted modifications increase continuously up to the kinematic limit $\xp=1$.

%Fig.~\ref{fig:jda} shows that within our framework, the
%modification to the centrality-dependent yields continues to increase
%up to the limit where the hard-scattered parton in the proton carries
%all of its longitudinal momentum, $\xp = 1$.

In addition to these future measurements, the high-\xp region can also
be accessed through measurements of hard-scattered yields at forward
rapidity (downstream of the projectile beam, $y \gtrsim +2$). Such
measurements have been previously performed in \sqsntwo \dau data by
PHENIX~\cite{Adler:2004eh,Adare:2011sc}, STAR~\cite{Abelev:2007nt} and
BRAHMS~\cite{Arsene:2004ux}. For our purposes, we focus on the PHENIX
measurement of di-hadron production~\cite{Adare:2011sc}, since
the centrality framework and selection is identical to that described in Sec.~\ref{sec:model}.
Particle production at moderate \pt in the forward
region predominantly arises from collisions with low-$x$ in the target
(Au) nucleus, $x_\mathrm{Au} \lesssim 10^{-2}$. The measurement shows
an increasing suppression in the quantity $J_{d+\mathrm{Au}}$, which is analogous
to \rdau but for azimuthally balanced di-hadron pairs instead of single
particles or jets, with decreasing $x_\mathrm{Au}$. This suppression has popularly been interpreted as arising from large, impact parameter dependent shadowing or parton saturation effects in the nuclear medium~\cite{Stasto:2011ru,Dumitru:2011vk}, or from impact parameter dependent energy loss~\cite{Kang:2011bp}. However, particle
production in this kinematic region arises from parton--parton
scatterings with large-\xp ($\xp > 0.1$) as well as
small $x_\mathrm{Au}$, and the results are therefore sensitive to the
physics of both, including the proton color fluctuation effects
described in this paper (We note that the importance of the high-\xp kinematics has been previously raised in a different context by the authors of Ref.~\cite{Kopeliovich:2005ym}).

In Ref.~\cite{Adare:2011sc}, centrality-dependent di-hadron yields were reported with two
selections on kinematics: one hadron at midrapidity ($|\eta| < 0.35$)
while the other hadron was at forward pseudorapidity ($3.1 < \eta < 3.8$),
and also with both hadrons in the forward pseudorapidity region. In general, the latter
selection is able to access lower $x_\mathrm{Au}$ and
higher \xp values. Results were reported within different ranges of the
leading and subleading hadron \pt, which were limited by kinematic
considerations to 0.5---0.75, 0.75---1.0 and 1.0---1.5 GeV/c. To apply
our framework to these results, di-hadron yields with a given
pseudorapidity and \pt selection need to be associated with a specific
value of \xp. {\sc PYTHIA}~\cite{Sjostrand:2006za} simulations were used to attempt to derive a mapping between the forward di-hadron kinematics and underlying value of \xp. However, in the simulation, string fragmentation processes without a tight association to the parton-parton kinematics contributed significantly to the production of particles in this kinematic range. Thus, following Ref.~\cite{Adare:2011sc}, we estimate the
mean \xp by assuming that the hadrons are the leading fragments of the
outgoing hard-scattered partons in a leading order $2\to2$ picture,
\begin{equation}
\xp = \frac{\left[ \left(p_{T,1}\exp(+\eta_{1}) +
p_{T,2}\exp(+\eta_{2})\right)/\left<z\right> \right]}{ \sqsn},
\label{eq:kinematics}
\end{equation}
where $(p_{T,1}, p_{T,2})$ and $(\eta_1, \eta_2)$ refer to
the leading and subleading hadron \pt and pseudorapidity, and the
typical fraction of the parton's \pt contained by the leading fragment
is taken to be $\left<z\right> = 0.6$.

The $J_{d+\mathrm{Au}}$ data are plotted as a function of estimated $\xp$, along with the results of our model calculation, in 
Fig.~\ref{fig:jda}. For
particle or jet production at fixed \pt, there is an anticorrelation
between \xp and $x_\mathrm{Au}$. However, the data in
Fig.~\ref{fig:jda} are compiled from different selections on $p_{T,1}$
and $p_{T,2}$, complicating this relationship. Thus, the data points
are plotted with different markers to distinguish three distinct
ranges of $x_\mathrm{Au}$.

In the most central (0--20\%) collisions, the $J_{d+\mathrm{Au}}$ data show a
substantially larger suppression than our calculation which
incorporates only \xp-dependent proton color fluctuation physics. For data
points arising from the smallest values of $x_\mathrm{Au} < 10^{-3}$,
this difference is more than a factor of five. In the most peripheral
collisions (60--88\%), the data are at or slightly below unity, while
the model predicts a substantial enhancement. Thus, our model
substantially overpredicts the $J_{d+\mathrm{Au}}$ values in events of all
centrality classes. 

%On the otherhand, the calculated and measured
%$J_{CP}$ are closer together. Nevertheless, our model clearly fails to
%describe the forward rapidity data by itself. 

%This is so even though this picture can describe the modifications of mid-rapidity jet
%production and analogous results in \ppb collisions at the LHC, which
%show that the modifications are a function only of \xp up to large
%rapidity ($y \sim +4$).

One possible explanation of this discrepancy is that assigning a definite value of \xp to this data is inappropriate.
The \dau di-hadron analysis assumes that the kinematics of the forward
rapidity measurement is related to the initial parton--parton
scattering kinematics \xp and $x_\mathrm{Au}$ in a well-defined way as
described in Eq.~\ref{eq:kinematics}. If such a correspondence does
not apply for particle production at such low \pt, then the above comparison is likely invalid.

%The \dau di-hadron analysis assumes that the kinematics of the forward
%rapidity measurement is related to the initial parton--parton
%scattering kinematics \xp and $x_\mathrm{Au}$ in a well-defined way as
%described in Eq.~\ref{eq:kinematics}. If such a correspondence does
%not apply for particle production at such low \pt, it is possible that no
%such unique $2 \rightarrow 2$ kinematics can be used.

%A physical picture based only on a shrinking proton is incomplete
%in this kinematic region for \dau collisions.     This is in striking contrast to the
%forward rapidity (up to $y \sim +4$) jet data in \ppb collisions at the LHC which have a centrality
%dependent modification that is nicely described as a function of only \xp~\cite{ATLAS:2014cpa}.
%It is notable that the \ppb data are for reconstructed jets at high \pt, which is in contrast to the \dau 
%data at low \pt.     Thus the differences may be attributable to the much lower momentum transfer scale $Q^{2}$
%for the RHIC data.   

%This is in striking contrast to the
%forward rapidity (up to $y \sim +4$) jet data in \ppb collisions at the LHC which have a centrality
%dependent modification that is nicely described as a function of only \xp~\cite{ATLAS:2014cpa}.

A second explanation is that additional physics effects must be accounted for. 
This is in striking contrast to the forward rapidity (up to $y \sim +4$) 
jet data in \ppb collisions at the LHC which have a centrality
dependent modification that is consistent with being only a function of \xp~\cite{ATLAS:2014cpa}.
However, the \ppb data are for reconstructed jets at much higher \pt values than those probed in the \dau di-hadron measurement. Thus the differences may be attributable to the much lower momentum transfer scale $Q^{2}$
for the RHIC data. 

An obvious candidate for an additional physics effect is the suppression of the parton density at low-$x_{\mathrm{Au}}$
described above. It notable that this 
effect would likely be very small for the \ppb data at the LHC. 
This suppression would decrease the $J_{d+\mathrm{Au}}$ relative
to the modifications arising from a high-\xp shrinking proton
effect. This could provide the additional suppression by a factor of
five in 0--20\% collisions needed to describe the data. A similar but
smaller low-$x_\mathrm{Au}$ suppression would also have to be present in
peripheral events to explain the data. While it is commonly believed
that peripheral \dau events are similar to \pp collisions and contain
no appreciable cold nuclear matter effects,
measurements of quarkonia
production~\cite{Adare:2010fn,McGlinchey:2012bp} found that even
60--88\% \dau collisions may have significant shadowing effects.

\begin{figure}
	\centering
	\includegraphics[width=0.98\columnwidth]{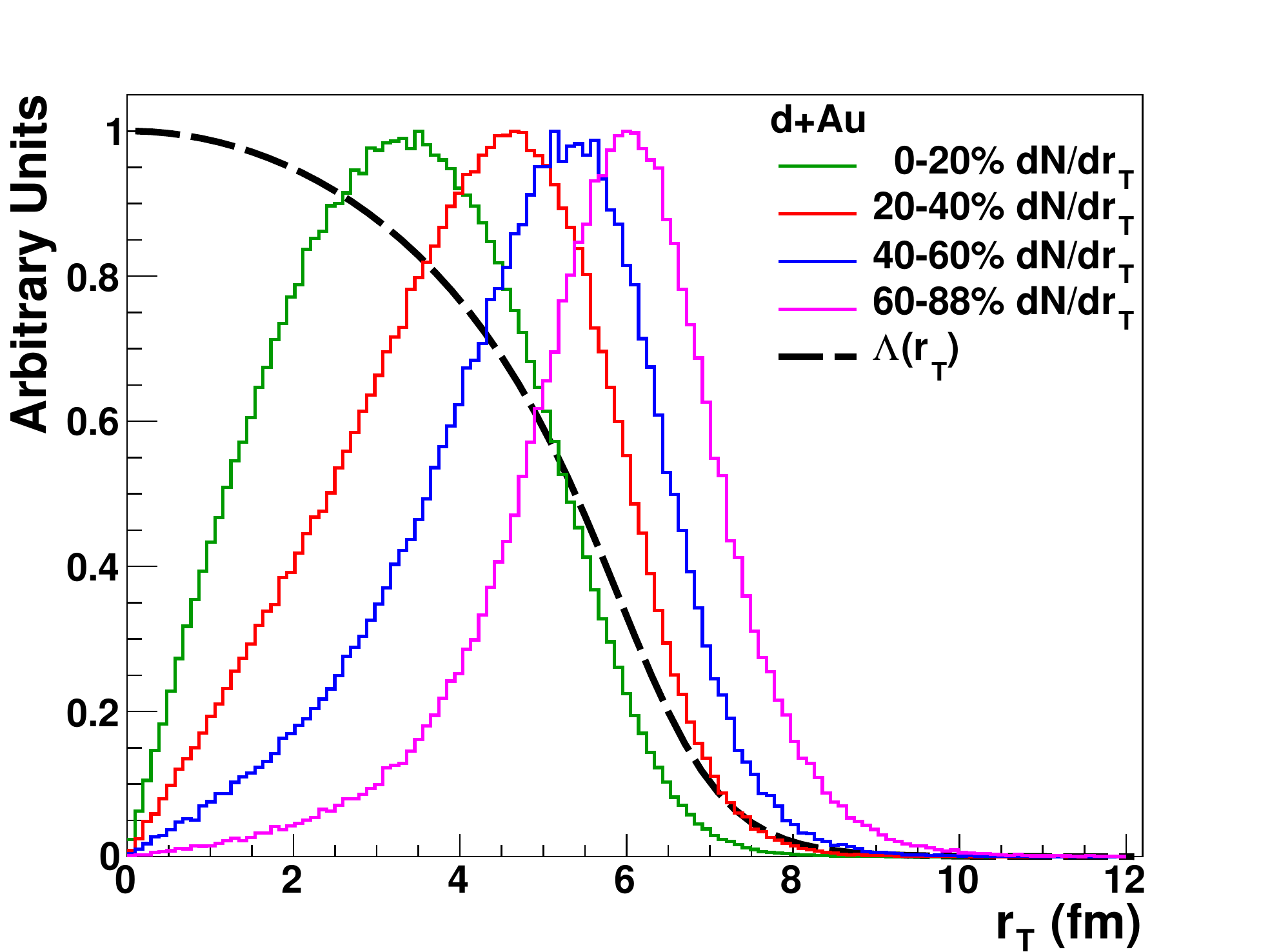}
	\caption{\label{fig:rt} Distributions of the radial impact position of nucleons ($r_T$) in \dau collisions for each of the four PHENIX centrality bins, reproduced from Ref.~\cite{Adare:2010fn}.  Also shown is the nuclear thickness as a function
    of the impact position, in arbitrary units.}
\end{figure}

Ref.~\cite{Adare:2010fn} calculates distributions of the projectile nucleon impact transverse 
radius, $r_T$, for each PHENIX \dau centrality selection, using 
a MC-Glauber model combined with a parameterization of the PHENIX 
centrality detector effects. Figure~\ref{fig:rt} shows these distributions 
in four centrality selections. The distributions in each selection overlap substantially with one another, 
and the mean $r_T$ between adjacent selections differs by typically less than 1 fm. 
Notably, the distribution in peripheral \dau collisions has a mean of $5.7$ fm (compared to $3.3$ fm in central collisions), and 
a long tail towards small $r_T$ values where the nucleus is thick.

%Figure~\ref{fig:rt} shows these results for four centrality selections.  
%The mean projectile nucleon $r_{T}$ for the most central and peripheral 
%\dau collisions were determined to be $3.3$ fm and $5.7$ fm, respectively.   
%Also shown in the Figure is the nuclear thickness as a function
%of the impact position.

We investigate if a small low-$x_\mathrm{Au}$ suppression in
peripheral events can, in addition to the high-\xp effects, describe
the data. To do this, we hypothesize that the centrality-dependent
low-$x_\mathrm{Au}$ suppression effect can be parameterized as a
linear function of the nuclear thickness,
\begin{equation}
J_{d+\mathrm{Au}}^{\ \prime} = J_{d+\mathrm{Au}}\left(1 - cT_{p+A}\right),
\label{eq:rdauprime}
\end{equation}

\noindent where $J_{d+\mathrm{Au}}$ contains the high-\xp effects, the parameter
$c$ encodes the strength of the centrality-dependent
low-$x_\mathrm{Au}$ suppression effects, and $J_{d+\mathrm{Au}}^{\ \prime}$ contains both
high-\xp and low-$x_\mathrm{Au}$ effects together. If $c$ is chosen to
produce the factor of $J_{d+\mathrm{Au}}^{\ \prime}/J_{d+\mathrm{Au}} = 0.2$ for $\xp > 0.4$
in central events, Eq.~\ref{eq:rdauprime} predicts a suppression of
$J_{d+\mathrm{Au}}^{\ \prime}/J_{d+\mathrm{Au}} = 0.6$ for peripheral events in this region
of \xp. 

Figure~\ref{fig:jda} shows the calculated $J_{d+\mathrm{Au}}^{\ \prime}$, which agrees well with the data in the region
$\xp > 0.4$ for each centrality selection. Thus, the
centrality-dependence of hadron production at forward rapidity in \dau
collisions is consistent with a combination of
effects from proton color fluctuations at high-\xp and an additional suppression from
low-$x_\mathrm{Au}$ effects linear in the nuclear thickness.

This analysis is not meant to replace a more detailed calculation of
impact parameter dependent low-$x_\mathrm{Au}$ effects, but instead is
meant to highlight that with a very strong physics effect in central
events, one may naturally expect a smaller, but still significant,
effect in peripheral events. Proton color fluctuations effects, which
may be logically expected from the midrapidity data, also play a role
in this kinematic region at low-$x_\mathrm{Au}$.

A better understanding of the relevant physics effects can be achieved
with a comparable di-hadron measurement in \pau collisions. In this system, cold nuclear matter shadowing effects would be very similar to those in \dau collisions. However, as we have argued in Sec~\ref{sec:results}, the high-\xp effects would be substantially larger, allowing for the separation of the two effects through the analysis of both collision systems.

\section{Summary}
\label{sec:summary}

This paper presents a simple geometric model of an \xp-dependent decrease in the interaction strength of the hard-scattered projectile nucleon in $p$/$d$/$^{3}$He$+$Au collisions. We implement this shrinking-nucleon picture using a MC-Glauber approach and calculate the resulting nuclear modification factors for centrality-dependent hard-scattering yields. After tuning, our one-parameter description of the \xp-dependence successfully describes the full \pt and centrality dependence of the measured \rdau and \rcp values for midrapidity jet production in \dau collisions at \sqsntwo. 

Having tuned our model to \dau collisions, we make quantitative predictions for \pt and centrality dependent nuclear modification factors in \pau and \heau collisions. We find that the one fewer (one additional) nucleon in \pau (\heau) collisions results in systematically larger (smaller) modifications arising from the shrinking of the projectile nucleon configurations with a large-\xp parton, relative to \dau collisions. This results in a clear ordering of the calculated \rpa and \rcp values based on projectile mass in the three collision systems. Recently collected data in these systems at RHIC can quantitatively test this picture. In particular, measurements of high-\pt jet or neutral pion production at midrapidity can distinguish the predicted $>50$\% differences in the most central to most peripheral \rcp ratio at $\xp \sim 0.3$.

We also explore the relevance of the shrinking nucleon configurations with a large-\xp parton for previous measurements of centrality-dependent di-hadron production in \dau collisions at forward rapidity. We find that while this large-\xp effect contributes to measurements in this kinematic regime, it alone does not provide a complete description of that data, with the additional modifications presumably arising from low-$x_\mathrm{Au}$ shadowing, saturation or energy loss effects in the nuclear medium. However, we demonstrate that the high-\xp effects must be taken into account to properly determine the impact parameter dependence of the low-$x_\mathrm{Au}$ or energy loss effects. Finally, we argue that an analogous measurement in \pau collisions, where the low-$x_\mathrm{Au}$ physics should be unchanged but the large-\xp physics more impactful, can further clarify the picture.

\begin{acknowledgments}

DM and JLN acknowledge funding from the Division of Nuclear
Physics of the U.S. Department of Energy under Grant
No. DE-FG02-00ER41152. DVP acknowledges funding from the U.S. Department of Energy under Contract No. DE-SC0012704. DVP also acknowledges Mark Strikman for useful discussions.

\end{acknowledgments}

\bibliography{main}

\end{document}